\documentclass{llncs}

\usepackage{amssymb}
\usepackage{graphicx}
\usepackage{enumitem}
\usepackage{multirow}
\usepackage{hhline}
\usepackage{url}
\usepackage{bbding}

\urldef{\mailsa}\path|konstantinos.chronis@gmail.com|
\urldef{\mailsb}\path|lucas.gren@cse.gu.se|
\urldef{\mailsc}\path||
\newcommand{\keywords}[1]{\par\addvspace\baselineskip
\noindent\keywordname\enspace\ignorespaces#1}

\begin{document}

\mainmatter  

\title{Agility Measurements Mismatch:\\
A Validation Study on Three Agile Team Assessments in Software Engineering \\}

\titlerunning{Agility Measurements Mismatch}

\author{Konstantinos Chronis\inst{1} \and Lucas Gren\inst{1,2}
}
\authorrunning{Konstantinos Chronis and Lucas Gren}

\institute{Chalmers and University of Gothenburg \\
SE-412 96 Gothenburg, Sweden\\ 
\and
University of S\~ao Paulo\\
S\~ao Paulo, Brazil 05508--090\\
\mailsa\\
\mailsb\\
\mailsc\\}

\toctitle{Lecture Notes in Computer Science}
\tocauthor{Authors' Instructions}
\maketitle

\begin{abstract}
Many tools have been created for measuring the agility of software teams, thus creating a saturation in the field. Three agile measurement tools were selected in order to validate whether they yield similar results. The surveys of the tools were given to teams in Company $A$ ($N=30$). The questions were grouped into agile practices which were checked for correlation in order to establish convergent validity. In addition, we checked whether the questions identified to be the same among the tools would be given the same replies by the respondents. We could not establish convergent validity since the correlations of the data gathered were very few and low. In addition, the questions which were identified to have the same meaning among the tools did not have the same answers from the respondents. We conclude that the area of measuring agility is still immature and more work needs to be done. Not all tools are applicable to every team but they should be selected on the basis of how a team has transitioned to agile. 
\end{abstract}

\section{Introduction}\label{sec:introduction}
Agile and plan-driven methodologies are the two dominant approaches in the software development. Although it has been almost 20 years since the former were introduced, the companies are quite reluctant in following them \cite{4599456}. 

Software development teams started adopting the most known agile methodologies, such as eXtreme Programming \cite{Beck:2004:EPE:1076267}, Feature Driven Development (FDD), \cite{Palmer:2001:PGF:600044}, Crystal \cite{Cockburn:2004:CCH:1406822}, 
Scrum \cite{scrum} and others. Most companies use a tailored methodology by following some of the aforementioned processes and practices which better suit their needs. Williams et al. \cite{williams2004toward} report that all XP practices are exercised rarely in their pure form, something on which Reifer \cite{Reifer} and Aveling \cite{aveling} also agree based on the results of their surveys, which showed that it is common for organizations to partially adopt XP. The most important issue that tends to be neglected though, is how well these methodologies are adopted.

According to Escobar-Sarmiento and Linares-Vasquez \cite{6427226}, the agile methodologies are easier to misunderstand. 
The previous statement is also supported by Taromirad and Ramsin \cite{cefam}, who argue that the agile software development methodologies are often applied to the wrong context. 
Sidky \cite{sidky_dissertation} defines the level of agility of a company as the amount of agile practices used. Considering this statement, a group that uses pair programming and collective code ownership at a very low level is more agile than a group which uses only pair programming but in a more efficient manner.

Williams et al. \cite{comparative_agility} pose the question ``\textit{How agile is agile enough}"? 
According to a survey conducted by Ambysoft \cite{ambysoft}, only 65\% of the agile companies that answered met the five agile criteria posed in the survey. 
Poonacha and Bhattacharya \cite{poonacha} mentioned that the different perceptions of agile practices when they are adopted are troublesome, since even people in the same team understand them differently, according to the result of a survey \cite{ambler}.

Since agile methodologies become more and more popular, there is a great need for developing a tool that can measure the level of agility in the organizations that have adopted them. 
For over a decade, researchers have been constantly coming up with models and frameworks in an effort to provide a solution. 

This case study comprises three tools which claim to measure the agility of software development teams using surveys. These tools are Perceptive Agile Measurement (PAM) \cite{pam}, Team Agility Assessment (TAA) \cite{Leffingwell}, Objectives Principles Strategies (OPS) \cite{sventha_dissertation}. The first one has been validated with a large sample of subjects, the second one is well-used by companies and the third one covers many agile practices. Since all three tools measure agility, convergent validity should be established among them to corroborate this. The surveys from the three tools were given to Company $A$ employees to answer. The analysis of the data was performed by grouping the survey questions in accordance to to agile practices. The correlation of these practices were the indications for establishing the convergent validity. Moreover, questions identified to have the same meaning among the tools should have the same answers from the respondents. The purpose of this study is to check whether these three tools will yield similar results.

\paragraph{Research Questions.}
\begin{enumerate}
	 \item Will PAM, TAA and OPS yield similar results?
	\begin{enumerate}[label={\roman*)},align=left]
  		\item Does convergent validity exist between the tools?
  		\item Will the questions that are exactly the same in the tools yield the same results?
	\end{enumerate}	
\end{enumerate}

\section{Case Study}
Any effort to see if the selected agility measurement tools are valid in what they do, would require to apply them to real software developments teams. According to Runeson and Host \cite{Runeson_Host}, a case study is ``a suitable research methodology for software engineering research since it studies contemporary phenomena in their natural context". As a result, a case study was selected as the most suitable means.

\subsection{Subject Selection}
Company $A$ is a United States company which operates in the Point Of Sales (POS) area. 
It has four teams with mixed members of developers and testers. The teams do not follow a specific agile methodology, but rather a tailored mix of the most famous ones which suits the needs of each team. Methodology A, as we can name it, embraces the practices from the various agile methodologies, some of them to a larger and some of them to a smaller extent. The analysis process created by Koch \cite{koch2005agile} was used for identifying these methodologies. The identification of the practices was done by observing and understanding how the teams work. 

\subsection{Data Collection}
In order to collect the data, an online survey was considered to be the best option, since it could be easily answered by each subject. 

For each of the tools, four surveys were created (one for each team). The data collection lasted about one month, while the surveys for each tool were conducted every ten days. None of the subjects was familiar with any of the tools.

Two subjects were requested to answer to the surveys first, in order to detect if there were any questions which could cause confusion, but also to see how much time is needed to complete a survey. Once the issues pointed out by the two subjects were fixed, the surveys were sent to the rest of the company's employees.

The links for the surveys were sent to the subjects via email, and they were asked to spend 15-20 minutes to reply to the survey. The employees who belonged to more than one team were asked a couple of days later to take the other survey in order to verify that their answers matched in both surveys. 


OPS agility measurements are based on three aspects: Adequacy, Capability and Effectiveness. Effectiveness measurement focuses on how well a team implements agile methodologies. Since the rest of the tools focus on the same thing, it was decided only to use the survey from Effectiveness and not to take into account the Adequacy and Capability aspects.

The surveys for PAM, TAA and OPS were answered on a Likert scale 1--7 (never having done what is asked in the question to always doing what is asked in the question).

The employees who were asked to answer to the surveys were all members of the software development teams, which consisted of software and QA engineers. All of the participating employees have been in the company for over a year and most of them have more than five years of work experience in an agile environment. Employees who had been working for less than six months in the company were not asked to participate, since it was considered that they were not fully aware of the company's procedures or that they were not familiar enough with them. Each participant replied to 176 questions in total. Initially, 34 surveys were expected to be filled in, but in the end, 30 of them were filled in, since some employees chose not to participate.

\subsection{Data Preparation}
\label{subsec:data_preparation}
All three tools have different amount of questions and cover different practices. For this reason, we preferred to do a grouping of the questions based on the practices/areas to which they belong.

\paragraph[TAA Areas]{Team Agility Assessment -- Areas.}
Team Agility Assessment (TAA) does not claim that it covers specific agile practices, but rather areas important for a team. It focuses on product ownership for Scrum teams but also on the release, iteration planning and tracking. The team factor plays a great role, as well as the development practices and the work environment. Automated testing and release planning are important here as well.

\paragraph[PAM Practices]{Perceptive Agile Measurement -- Practices.}
The Perceptive Agile Measurement (PAM) tool focuses on the iterations during software development, but also on the stand-up meetings for the team members, their collocation and the retrospectives they have. The access to customers and their acceptance criteria have a high importance as well. Finally, the continuous integration and the automated unit testing are considered crucial in order to be agile.

\paragraph[OPS Practices]{Objectives, Principles, Strategies (OPS) -- Practices.}
Objectives, Principles, Strategies (OPS) Framework is the successor of the Objectives, Principles, Practices (OPP) Framework \cite{opp}. OPP identified 27 practices as implementations of the principles which later on were transformed into 17 strategies.

\paragraph[Tool Practices]{Practices Covered Among The Tools.}
\label{subsubsec:practices_among_tools}
We have abstracted some of the OPP practices to OPS strategies in order to avoid repeating the mapping of the questions. The connection between the practices and the strategies is done based on the questions of each tool.

\paragraph{Mapping of questions among tools.}
\label{subsubsec:mapping}

PAM has its questions divided on the basis of agile practices, while on the other hand, TAA has divided them based on areas considered important. Although all practices/areas from PAM and TAA are mapped onto OPP and OPS, not all of their questions are under OPP practices or OPS strategies. This can be explained due to the different perception/angle that the creators of the tools have and what is considered important for an organization/team to be agile.

\subsection{Data Analysis}

The data gathered from the surveys were grouped on the basis of the practices covered by the OPP, and as a consequence, the OPS.

\paragraph{Convergent Validity Analysis.}
\label{subsubsec:convergent_validity_analysis}
Since all the tools claim to be measuring agility and under the condition that convergent validity exists among them, then, by definition, they should yield similar results. 

In similar studies \cite{jalali_angelis,Delestras2013}, the correlation analysis was selected as the best way to check similar tools and this was followed here as well. 
We decided to use the practices covered by each tool and see if they correlate with the same practices from the other two tools. The idea is based on the \textit{multitrait-multimethod matrix}, presented by Campbell and Fiske \cite{campbell1959}. The matrix is the most commonly used way for providing construct validity. 

In order to select which correlation analysis method to choose from, the data were checked if they had normal distribution by using the Shapiro-Wilk test which is the most powerful normality test, according to a recent paper published by Razali and Wah \cite{Razali}. The chosen alpha level was 0.05, as it is the most common one.


Out of the 42 normality checks (three for each of the 14 practices), only 17 concluded that the data are normally distributed. The low level of normally distributed data gave a strong indication that Spearman's rank correlation coefficient, which is more adequate for non-parametric data, was more appropriate to use, rather than the Pearson's product-moment correlation.

In order to use the Spearman's rank correlation coefficient, a monotonic relationship between two variables is required. In order to check for the monotonicity, plots were drawn between the results of each tool for all 14 practices. The plots surprisingly showed that only eight out of 42 were monotonic, which indicates no correlation what-so-ever. 

\paragraph{Direct Match Questions Analysis.}
\label{subsubsec:direct_match_analysis}
We want to find which questions are the same among the tools. In order to achieve this, the mapping described in subsection~\ref{subsubsec:mapping} was used. Afterward, the questions were checked one by one to identify the ones which had the same meaning. When we finalized the groups of questions which were the same, we requested from the same employees who were taking the pilot surveys to verify if they believed the groups were correctly formed. Their answer was affirmative, so we continued by checking if the answers of the subjects were the same. Surprisingly, OPS--TAA have 20 questions with the same meaning, while OPS--PAM and TAA--PAM only four and three respectively. 


Out of the 35 normality checks (two for each group and three for one group), only 2 concluded that the data are normally distributed. Since the samples are also independent (they do not affect one another), there is a strong indication that the Mann–Whitney U test is appropriate. For the group \textit{Smaller And Frequent Product Releases}, we used the Kruskal--Wallis one-way analysis of variance method, which is the respective statistical method for more than two groups.

The hypothesis in both cases was:
\begin{itemize}[label={}]
	\item $H_0$: \textit{There is no difference between the groups of the same questions}
	\item $H_1$: \textit{There is a difference between the groups of the same questions}
\end{itemize}


\section{Results}

\subsection{Correlations}
As it was previously stated, only eight out of 42 plots were monotonic. The more interesting than the correlations result is the non-existence of monotonicity in the other 34 relationships, which leads us to the conclusion that there is little convergence among the tools. This is surprising because tools claiming to measure the same thing should converge.  

\subsection[Direct Match Results]{Direct Match Questions Results}
\label{sec:direct_match_results}

The groups of direct match questions showed some unexpected results. 
Questions which are considered to have the same meaning should yield the same results, which was not the case for any of the question groups, apart from one group concerning the \textit{Software Configuration Management}. 
On the other hand, the \textit{Product Backlog} practice had the lowest score with only six respondents giving the same answer. The maximum difference in answers was up to two Likert-scale points.

As far as the results from the Mann–-Whitney U test and Kruskal–-Wallis one-way analysis of variance are concerned, the p-values from the majority of the groups are more than the alpha level of 0.05. As a result, we cannot reject the $H_0$ hypothesis. 
Such practices are  \textit{Iteration Progress Tracking and Reporting - group \#2}, \textit{High-Bandwidth Communication} and others.
On the other hand, the $p$-value of group \textit{Software Configuration Management} cannot be computed, since all the answers are the same, while for other groups the $p$-value is below the alpha level which means that the $H_0$ hypothesis can be rejected. Such practices are \textit{Continuous Integration - group \#2}, \textit{Iteration Progress Tracking and Reporting - group \#4} and others.


\section{Discussion}\label{sec:discussion}

\subsection{Will PAM, TAA and OPS yield similar results?}
The plots drawn by the data gathered showed an unexpected and interesting result. Not only do the tools lack a correlation, but they do not even have a monotonic relationship when compared to each other for the agile practices covered, resulting in absence of convergent validity. This could indicate two things; the absence of monotonicity and the negative or very low correlations show that the questions used by the tools in order to cover an agile practice do it differently as well as that PAM, TAA and OPS measure the agility of software development teams in their own unique way. 




Almost all groups had different responses to the same questions. 
With regards to the research question \textbf{\textit{``Does convergent validity exist among the tools?"}}, we showed that convergent validity could not be established due to the low (if existing) correlations among the tools. Concerning the research question \textbf{\textit{``Will the questions that are exactly the same among the tools yield the same results?"}}, we saw that a considerable amount of respondents' answers were different.


The reasons for this somewhat unexpected results are explained in the following paragraphs.

\paragraph{Few or no questions for measuring a practice.}
A reason for not being able to calculate the correlation of the tools is that they cover slightly or even not at all some of the practices. An example of this is the \textit{Smaller and Frequent Product Releases} practice. OPS includes four questions, while on the other hand, PAM and TAA have a single question each. Furthermore, \textit{Appropriate Distribution of Expertise} is not covered at all by PAM. In case the single question gets a low score, this will affect how effectively the tool will measure an agile practice. On the contrary, multiple questions can better cover the practice by examining more factors that affect it.

\paragraph{The same practice is measured differently.}
Something interesting that came up during the data analysis was that although the tools cover the same practices, they do it in different ways, leading to different results. An example of this is the practice of \textit{Refactoring}. PAM checks whether there are enough unit tests and automated system tests to allow the safe code refactoring. In case the course unit/system tests are not developed by a team, the respondents will give low scores to the question, as the team members in Company $A$ did. Nevertheless, this does not mean that the team never refactors the software or does it with bad results. All teams in Company $A$ choose to refactor when it adds value to the system, but the level of unit tests is very low and they exist only for specific teams. On the other hand, TAA and OPS check how often the teams refactor, among other aspects.

\paragraph{The same practice is measured in opposite questions.}
The \textit{Continuous Integration} practice has a unique paradox among TAA, PAM and OPS. The first two tools include a question about the members of the team having synchronized to the latest code, while OPS checks for the exact opposite. According to Soundararajan \cite{sventha_dissertation}, it is preferable for the teams not to share the same code in order to measure the practice.

\paragraph{Questions phrasing.}
Although the tools might cover the same areas for each practice, the results could differ because of how a question is structured. An example of this is the \textit{Test Driven Development} practice. Both TAA and PAM ask about automated code coverage, while OPS just asks about the existence of code coverage. Furthermore, TAA focuses on 100\% automation, while PAM does not. Thus, if a team has code coverage but it is not automated, then the score of the respective question should be low. In case of TAA, if the code coverage is not fully automated, its score should be even lower. It is evident that the abstraction level of a question has a great impact. The more specific it is, the more a reply to it will differ, resulting in possible low scores.


\paragraph{Better understanding of agile concepts.}
In pre-post studies there is a possibility of the subjects becoming more aware of a problem in the second test due to the first test \cite{Campbell_Stanley}. Although the \textit{testing} threat, as it is called, does not directly apply here, the similar surveys on consecutive weeks could have enabled the respondents to take a deeper look into the agile concepts, resulting in better understanding of them, and consequently, providing different answers to the surveys' questions. 

\paragraph{How people perceive agility.}
Although the concept of agility is not new, people do not seem to fully understand it, as Conboy and Wang \cite{Wang_Conboy} also mention. This is actually the reason behind the existence of so many tools in the field which are trying to measure how agile the teams are or the methodologies used. The teams implement agile methodologies differently and researchers create different measurement tools. There are numerous definitions of what agility is \cite{Kidd,Kara,Ramesh,agile_manufacturing}, and each of the tool creators adopt or adapt the tools to match their needs. Their only common basis is the agile manifesto and its twelve principles \cite{beck2001agile}, which are (and should be considered as) a compass for the agile practitioners. Nevertheless, they are not enough and this resulted in the saturation of the field. Moreover, Conboy and Fitzgerald \cite{conboy_fitzgerald} state that the agile manifesto principles do not provide practical understanding of the concept of agility. Consequently, all the reasons behind the current survey results are driven by the way in which tool creators and tool users perceive agility.

The questions in the surveys were all based on how their creators perceived the agile concept which is quite vague, as Tsourveloudis and Valavanis \cite{tsourveloudis} have pointed out. 
None of the Soundararajan \cite{sventha_dissertation}, So and Scholl \cite{pam}, Leffingwell \cite{Leffingwell} claimed, of course, to have created the most complete measurement tool, but still, this leads to the oxymoron that the tools created by specialists to measure the agility of software development teams actually do it differently and without providing substantial solution to the problem. On the contrary, this leads to more confusion for the agile practitioners.

Considering that the researchers and specialists in the agile field perceive the concept of agility differently, it would be naive to say that the teams do not do the same. The answers to surveys are subjective and people reply to them depending on how they understand them. Ambler \cite{ambler} stated the following: ``I suspect that developers and management have different criteria for what it means to be agile''. This is also corroborated by the fact that, although a team works in the same room and follows the same processes for weeks, it is rather unlikely that its members will have the same understanding of what a retrospection or a releasing planning meeting means to them, a statement which is also supported by Murphy et al. \cite{Williams_Microsoft}. 

\section{Threats to Validity}

\subsection{Construct Validity}
We consider that the construct validity concerning the surveys given to the subjects was already handled by the creators of the tools which were used. Our own construct validity lies in establishing the convergent validity. The small sample of subjects was the biggest threat in establishing convergent validity, making the results very specific to Company $A$ itself. Future work on this topic should be performed at other companies to mitigate this threat. In order to avoid mono-method bias, some employees were asked to fill in the surveys first in order to detect any possible issues. All the subjects were promised to remain anonymous, resulting in mitigating the evaluation apprehension \cite{wohlin2012expse}. 

\subsection{Internal Validity}
The creators of PAM, TAA and OPS have already tried to mitigate internal validity when creating their tools. Yet, there are still some aspects of internal validity, such as selection bias maturation and testing effect. With regard to maturation, this concerns the fatigue and boredom of the respondents. Although the surveys were small in size and did not require more than 15-20 minutes each, still the similar and possibly repetitive questions on the topic could cause fatigue and boredom to the subjects. This could result in the participants giving random answers to the survey questions. The mitigation for this threat was to separate the surveys and conduct them during three different periods. In addition, the respondents could stop the survey at any point and continue whenever they wanted. As far as the testing effect is concerned, this threat could not be mitigated. The testing effect threat applies to pre-post design studies only, but due to the same topic of the surveys, the subjects were to some extent more aware of what questions to expect in the second and third survey. Finally, selection could also not be mitigated, since the case study focused on a specific company only.

\subsection{Conclusion Validity}
Although the questions of the surveys have been carefully phrased by their creators, still there may be uncertainty about them. In order to mitigate this, for each survey a pilot one was conducted to spot any questions which would be difficult to understand. In addition, the participants could ask the first author about any issue they had concerning the survey questions. Finally, the statistical tests were run only for the data that satisfied the prerequisites, with the aim to mitigate the possibility of incorrect results. 

\subsection{External Validity}
This case study was conducted in collaboration with one company and 30 subjects only. Consequently, it is hard to generalize the outcomes. Nevertheless, we believe that any researcher replicating the case study in another organization with teams which follow the same agile practices as those used in Company $A$ would get similar results.

\subsection{Reliability}
To enable other researchers to conduct a similar study, the steps followed have been described and the reasons for the decisions made have been explained. Furthermore, all the data exist in digital format which can be provided to anyone who wants to review them. The presentation of the findings could probably be a threat to validity because of the first author's experience at the company. In order to mitigate this, the findings were discussed with a Company $A$ employee who did not participate in the case study.

\section{Conclusions and Future Work}

\subsection{Conclusions}

This paper contributes to the area of measuring the agility of software developments teams. This contribution can be useful for the research community, but mostly for practitioners. We provided some evidence that tools claiming to measure agility do not yield similar results. The expertise of the tool creators is unquestionable, but nevertheless, their perception of agility and their personal experience have led them to create a tool in the way they consider appropriate. A measurement tool which satisfies the needs of one team may not be suitable for other teams. This derives not only from the team's needs but also from the way it transitioned to agile. 
Companies need a tool to measure agility in order to identify their mistakes and correct them with the total purpose to produce good quality software for their customers. There is still work to be done in order to find a universal tool for measuring agility, and such a tool should be scientifically validated before it is used. 

\subsection{Future Work}

It would be interesting to see the results of a study that would be conducted at more companies, in order to compare them to the results of the present study. In addition, another way of forming the data samples could indicate different results, which is worth looking into. Moreover, future work in the field could check for establishing convergent validity among other agility measurement tools, combine them, validate them, and finally, only use them where their output is relevant in context.

\bibliographystyle{splncs}
\bibliography{references}
\newpage

\end{document}